\documentclass[doublespacing]{elsart} 
\usepackage{graphicx,amssymb}

\newlength{\figurewidth}
\begin{document}
\begin{frontmatter}

\title{First measurement of the \\ \nuc{14}{N}(p,$\gamma$)\nuc{15}{O} cross section down to 70\,keV	}

\author[infnge]{A.\,Lemut}, 
\author[infnpd]{D.\,Bemmerer},
\author[infnge]{F.\,Confortola}, 
\author[infnmi]{R.\,Bonetti}, 
\author[infnpd]{C.\,Broggini\corauthref{broggini}}, 
	\corauth[broggini]{Corresponding Author, {\tt broggini@pd.infn.it}, phone: {\tt +39 049 827 7123}, fax: {\tt +39 049 827 7145}.}
\author[infnge]{P.\,Corvisiero},
\author[infnge]{H.\,Costantini},
\author[lisboa]{J.\,Cruz},
\author[lngs]{A.\,Formicola},
\author[atomki]{Zs.\,F\"ul\"op},
\author[infnto]{G.\,Gervino},
\author[infnmi]{A.\,Guglielmetti},
\author[lngs]{C.\,Gustavino},
\author[atomki]{Gy.\,Gy\"urky},
\author[unina]{G.\,Imbriani},
\author[lisboa]{A.P.\,Jesus},
\author[lngs]{M.\,Junker},
\author[unina]{B.\,Limata},
\author[infnpd]{R.\,Menegazzo},
\author[infnge]{P.\,Prati},
\author[unina]{V.\,Roca},
\author[caserta]{D.\,Rogalla},
\author[bochum]{C.\,Rolfs},
\author[unina]{M.\,Romano},
\author[infnpd]{C.\,Rossi Alvarez},
\author[bochum]{F.\,Sch\"umann},
\author[atomki]{E.\,Somorjai},
\author[teramo]{O.\,Straniero},
\author[bochum]{F.\,Strieder},
\author[caserta]{F.\,Terrasi},
\author[bochum]{H.P.\,Trautvetter} \\
\collab{The LUNA Collaboration}
%
% Institutes
	\address[infnge]{Dipartimento di Fisica, Universit\`a di Genova, and INFN, Genova, Italy }
	\address[infnpd]{Istituto Nazionale di Fisica Nucleare (INFN), Sezione di Padova, via Marzolo 8, 35131 Padova, Italy}
	\address[infnmi]{Istituto di Fisica, Universit\`a di Milano, and INFN, Milano, Italy}
	\address[lisboa]{Centro de Fisica Nuclear da Universidade de Lisboa, Lisboa, Portugal}
	\address[lngs]{INFN, Laboratori Nazionali del Gran Sasso, Assergi, Italy}
	\address[atomki]{ATOMKI, Debrecen, Hungary}
	\address[infnto]{Dipartimento di Fisica Sperimentale, Universit\`a di Torino, and INFN, Torino, Italy}
	\address[unina]{Dipartimento di Scienze Fisiche, Universit\`a di Napoli "Federico II", and INFN, Napoli, Italy}
	\address[caserta]{Seconda Universit\`a di Napoli, Caserta, and INFN, Sezione di Napoli, Napoli, Italy}
	\address[bochum]{Institut f\"ur Experimentalphysik III, Ruhr-Universit\"at Bochum, Bochum, Germany}
	\address[teramo]{Osservatorio Astronomico di Collurania, Teramo, and INFN, Sezione di Napoli, Napoli, Italy}

%%%%%%%%%%%%%%%%%%%%%%%%%%%%%%%%%%%%%
\begin{abstract}
In stars with temperatures above 20$\cdot$10$^6$\,K, hydrogen burning is dominated by the CNO cycle. Its rate is determined by the slowest process, the \nuc{14}{N}(p,$\gamma$)\nuc{15}{O} reaction. 
Deep underground in Italy's Gran Sasso laboratory, at the LUNA 400\,kV accelerator, the cross section of this reaction has been measured at energies much lower than ever achieved before. 
Using a windowless gas target and a 4$\pi$ BGO summing detector, direct cross section data has been obtained down to 70\,keV, reaching a value of 0.24 picobarn. The Gamow peak has been covered by experimental data for several scenarios of stable and explosive hydrogen burning.
In addition, the strength of the 259\,keV resonance has been remeasured.
The thermonuclear reaction rate has been calculated for temperatures 90\,--\,300\,$\cdot$10$^6$\,K, for the first time with negligible impact from extrapolations.
\end{abstract}
%%%%%%%%%%%%%%%%%%%%%%%%%%%%%%%%%%%%%

\end{frontmatter}

%%%%%%%%%%%%%%%%%%%%%%%%%%%%%%%%%%%%%

% =========================================================
\section{Introduction}
% =========================================================

Charged particle induced thermonuclear reactions in a star take place in a narrow energy window called the Gamow peak, far below the Coulomb barrier. Their cross section $\sigma(E)$ drops steeply with decreasing energy and can be parameterized as \cite{Rolfs88}:
\begin{equation} \label{eq:S-factor-def}
\sigma(E) = \frac{S(E)}{E} e^{-2 \pi \eta} 
\end{equation}
where $S(E)$ is the astrophysical S-factor, and $\eta$\,$\propto$\,$E^{-0.5}$ is the Sommerfeld parameter\footnote{In the present work, $E$ denotes the energy in the center of mass system, and $E_{\rm p}$ is the projectile energy in the laboratory system.}. 
The very low value of $\sigma(E)$ at stellar Gamow peak energies prevents a measurement in a laboratory at the surface of the Earth, where the signal to background ratio is too small because of cosmic ray interactions. 
In order to overcome this difficulty, a 50\,kV \cite{Greife94-NIMA} and a 400\,kV \cite{Formicola03-NIMA} accelerator have been installed in the Laboratory for Underground Nuclear Astrophysics (LUNA), deep underground in Italy's Gran Sasso laboratory. 
At LUNA, two reactions from the hydrogen burning p--p chain have been studied for the first time directly at their respective solar Gamow peak: \nuc{3}{He}(\nuc{3}{He},2p)\nuc{4}{He} \cite{Junker98-PRC,Bonetti99-PRL} and \nuc{2}{H}(p,$\gamma$)\nuc{3}{He} \cite{Casella02-NPA}.\par

The second process of stable hydrogen burning in stars, the carbon--nitrogen--oxygen (CNO) cycle, dominates for 20\,$\leq$\,$T_6$\,$\leq$\,130 ($T_6$: stellar temperature in 10$^6$\,K), corresponding to Gamow peak energies of 30\,--\,110\,keV. The slowest process, the \nuc{14}{N}(p,$\gamma$)\nuc{15}{O} reaction, has been studied previously \cite[and references therein]{Schroeder87-NPA}, but only one work reported data at these low energies
\cite{Lamb57-PR}.
Reaction rate compilations \cite{CF88-ADNDT,Adelberger98-RMP,NACRE99-NPA} have been based mainly on ref. \cite{Schroeder87-NPA}, with data down to $E$\,=\,181\,keV, and on ref. \cite{Lamb57-PR}. Subsequently, the extrapolation to solar energy of ref. \cite{Schroeder87-NPA} for capture to the ground state in \nuc{15}{O} has been challenged by several works, on theoretical \cite{Angulo01-NPA} and indirect \cite{Bertone01-PRL,Mukhamedzhanov03-PRC,Yamada04-PLB} grounds. 
These considerations \cite{Angulo01-NPA,Bertone01-PRL,Mukhamedzhanov03-PRC,Yamada04-PLB} were then experimentally confirmed at $E$\,= 119\,--\,367\,keV \cite{Formicola04-PLB,Imbriani05-EPJA}, indicating an extrapolated S-factor at solar energy that is a factor 2 smaller than the value adopted in the compilations.
Recently, an independent study reported cross section data for $E$\,=\,134\,--\,482\,keV, with identical conclusions for the total extrapolated S-factor \cite{Runkle05-PRL}. \par

In the present work, the results of a novel \nuc{14}{N}(p,$\gamma$)\nuc{15}{O} experiment are presented. 
The previous studies (table \ref{Motivation-table}) relied on extrapolations in order to predict the S-factor at astrophysical energies \cite{Schroeder87-NPA,Angulo01-NPA,Bertone01-PRL,Mukhamedzhanov03-PRC,Yamada04-PLB,Formicola04-PLB,Imbriani05-EPJA,Runkle05-PRL} or cover only a narrow energy range \cite{Lamb57-PR}.  
The present work follows a different approach. Cross sections are measured directly, with high statistics and over a wide, astrophysically relevant energy range. \par

% =========================================================
\section{Experiment}
% =========================================================

Equation \ref{eq:S-factor-def} predicts a factor 200 drop in the yield from $E$\,=\,119\,keV \cite{Formicola04-PLB,Imbriani05-EPJA} to 70\,keV (the aim of the present study), so a new setup with a thin, pure gas target and a high-efficiency annular BGO detector had to be developed \cite{Casella02-NIMA}. In the experiment, the proton beam with $E_{\rm p}$\,=\,80\,--\,250\,keV and 0.3\,mA typical current from the LUNA 400\,kV accelerator \cite{Formicola03-NIMA} passed several apertures (final aperture: 40\,mm long, 7\,mm diameter) and entered a windowless gas target.
The 12\,cm long target cell is placed inside the borehole (6\,cm diameter) of a 28\,cm long annular bismuth germanate (BGO) crystal in 4$\pi$ geometry, ensuring $\approx$70\,\% peak detection efficiency for 7\,MeV $\gamma$ rays \cite{Casella02-NIMA}. The beam is stopped on a copper disk that also serves as the hot side of a calorimeter with constant temperature gradient \cite{Casella02-NIMA}. \par 

The beam energy has been calibrated previously for $E_{\rm p}$\,=\,130\,--\,400\,keV \cite{Formicola03-NIMA}, and the calibration is extrapolated down to $E_{\rm p}$\,=\,80\,keV, resulting in 0.3\,keV uncertainty. The target gas was 1\,mbar nitrogen (corresponding to typically 10\,keV target thickness), 99.9995\,\% chemically pure. The pressure was monitored with a capacitance gauge and kept constant to better than 0.25\,\% by a feedback system. The beam heating effect \cite{Goerres80-NIM} reduced the target density by up to 15\,\% with incident ion beam. Using the resonance scan technique \cite{Rolfs88,Goerres80-NIM}, this correction was measured for the present setup, prior to the actual experiment. \par

% ============ Spectrum.eps

\begin{figure*}[p]
 \centering 
 \includegraphics[width=\figurewidth]{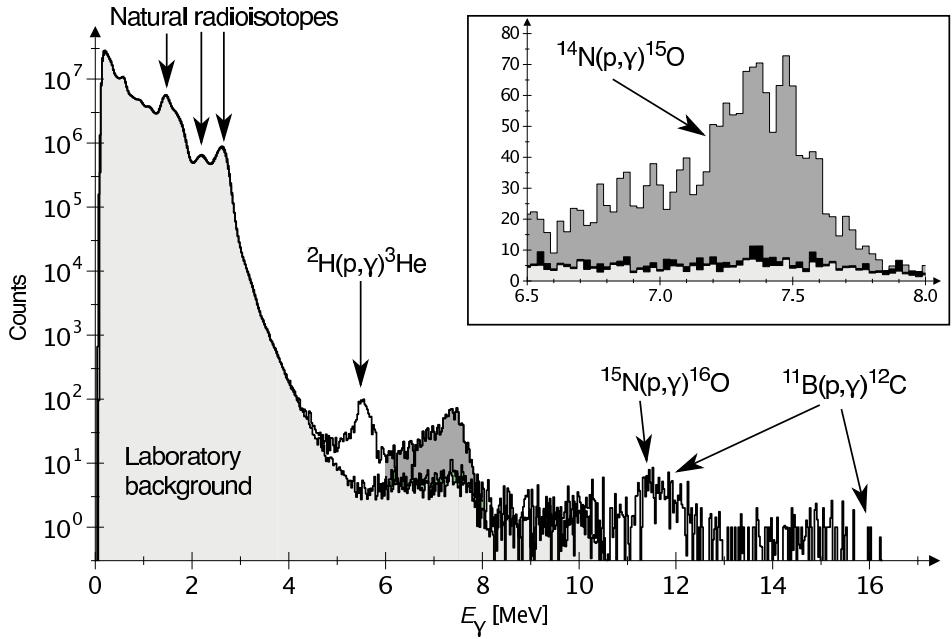}
 \caption{Typical $\gamma$ ray spectrum in the 4$\pi$ BGO detector, at $E$\,=\,90\,keV, livetime 11.4 days, accumulated charge 231 C. The counts from the reaction to be studied are shaded in dark gray. The laboratory background \cite{Bemmerer05-EPJA}, normalized to equal livetime, is shaded in light gray. The most important components of background induced by the ion beam are indicated. Inset: Region of interest (ROI) for the present study, with the beam induced background in the ROI indicated by the black filled area (see text).}
 \label{Spectrum.eps}
\end{figure*}

Because of the high absolute detection efficiency and of the near 4$\pi$ geometry of the BGO detector, $\gamma$ rays emitted in a cascade are with high probability summed into a peak at $E_\gamma = Q + E$ ($Q$\,=\,7.297\,MeV is the $Q$ value for the \nuc{14}{N}(p,$\gamma$)\nuc{15}{O} reaction) in the spectrum (fig. \ref{Spectrum.eps}). 
Next to the summing peak are unresolved lines at 6.172 and 6.792\,MeV, the energies of two secondary $\gamma$ rays. A peak at 5.6\,MeV results from the \nuc{2}{H}(p,$\gamma$)\nuc{3}{He} beam induced background reaction \cite{Bemmerer05-EPJA}. The broad structure at 12\,MeV is caused by the \nuc{15}{N}(p,$\gamma$)\nuc{16}{O} reaction (the target gas has natural isotopic composition, 0.4\,\% \nuc{15}{N}), with a contribution from the \nuc{11}{B}(p,$\gamma$)\nuc{12}{C} reaction. The latter reaction also gives $\gamma$ rays at 16\,MeV. The beam induced background in the region of interest (ROI) results from the Compton continuum due to high energy $\gamma$ rays (from \nuc{15}{N}(p,$\gamma$)\nuc{16}{O} and \nuc{11}{B}(p,$\gamma$)\nuc{12}{C}), and from the \nuc{13}{C}(p,$\gamma$)\nuc{14}{N} reaction ($Q$\,=\,7.551\,MeV). 
The number of events from \nuc{13}{C}(p,$\gamma$)\nuc{14}{N} ($E_\gamma$\,$\approx$\,7.7\,MeV) 
% inside the ROI 
has been evaluated by taking monitor $\gamma$ spectra with helium instead of nitrogen as target gas. The bulk of the \nuc{13}{C} 
% impurity 
(due to hydrocarbons from pump oil) seen by the beam has been localized at the beam stop \cite{Bemmerer05-EPJA}. The \nuc{11}{B} has been localized on the final collimator \cite{Bemmerer05-EPJA}, and it is 
due to impurities in the 
collimator material. % new daniel 03.02.2006
% material itself.
The laboratory background in the ROI is due to (n,$\gamma$) reactions (caused by ($\alpha$,n) neutrons from the laboratory background) mainly in the stainless steel target chamber, and it has a constant counting rate \cite{Bemmerer05-EPJA}. The total $\gamma$ ray background amounts to less than 20\,\% of the counts observed in the ROI for each run, except for the two lowest energy points: At $E$\,=\,80\,keV, the background is 45\,\% of the collected events. At E\,=\,70\,keV, with 53 days of running time, there are 11 counts/day from the reaction, 22 counts/day from laboratory background and 1 count/day from beam induced background. \par

The $\gamma$ ray detection efficiency depends on the measured, energy dependent branching ratios \cite{Imbriani05-EPJA}, and for $E$\,$<$\,119\,keV, where there is no data, on R-matrix extrapolations \cite{Imbriani05-EPJA}. However, this dependence is weakened \cite{Bemmerer05-EPJAdirect} by the flatness of the efficiency curve \cite{Casella02-NIMA} and the dominance of the 6.792\,MeV transition \cite{Schroeder87-NPA,Angulo01-NPA,Mukhamedzhanov03-PRC,Formicola04-PLB,Imbriani05-EPJA,Runkle05-PRL,Nelson03-PRC}. As a result, the branching ratios contribute 3\,\% systematic uncertainty in the detection efficiency, with an additional 1\,\% from the modeling of the detector \cite{Casella02-NIMA} and 1.5\,\% from the absolute efficiency calibration, giving a total of 3.5\,\%. The angular distributions for primary and secondary $\gamma$ rays from the five strongest transitions have been measured previously above and below the 259\,keV resonance \cite{Imbriani05-EPJA}, and isotropy was found for all $\gamma$ rays except for the primary from capture into the state at 6.792\,MeV. Since both the 6.792\,MeV secondary $\gamma$ ray and the sum peak are fully included in the ROI, the detection efficiency is unaffected \cite{Bemmerer05-EPJAdirect}. Details on experiment and analysis can be found elsewhere \cite{Bemmerer04-Diss,Lemut05-Diss} and will be published separately. \par

% =========================================================
\section{Results}
\begin{figure*}[p]
 \centering 
 \includegraphics[width=\figurewidth]{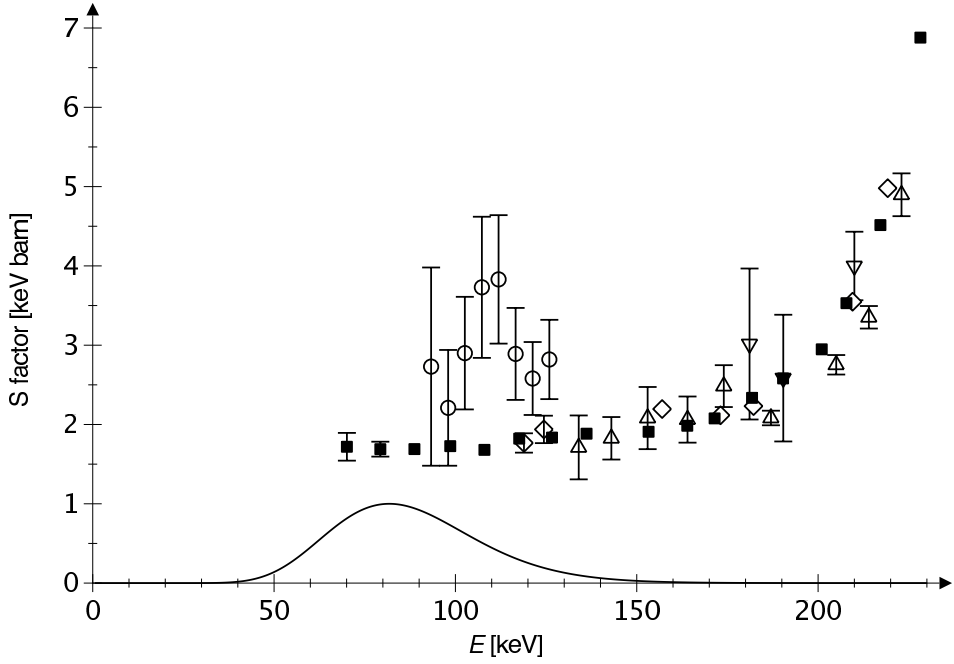}%
 \caption{Astrophysical S-factor for the \nuc{14}{N}(p,$\gamma$)\nuc{15}{O} reaction from the present work (filled squares) and from previous studies: circles \cite{Lamb57-PR}, inverted triangles \cite{Schroeder87-NPA}, diamonds \cite{Formicola04-PLB,Imbriani05-EPJA}, triangles \cite{Runkle05-PRL}. Error bars ($\pm$$1\sigma$ statistical uncertainty) are only shown where they are larger than the symbols used. The Gamow peak for $T_6$\,=\,80 is also shown. The systematic uncertainties are given in the text and in table \ref{Motivation-table}.}
 \label{Sfactor.eps}
\end{figure*}

% =========================================================
%
The present S-factor data (fig. \ref{Sfactor.eps}) reach a much lower energy than the previous direct experiments \cite{Schroeder87-NPA,Lamb57-PR,Formicola04-PLB,Imbriani05-EPJA,Runkle05-PRL}, while overlapping over a wide energy range. The statistical uncertainties are lower, and the systematic uncertainties are comparable or lower (table \ref{Motivation-table}). The systematic uncertainty for the lowest energy point in the present study is 
7\,\%, dominated by beam energy calibration (5\,\%), detection efficiency (3.5\,\%) and effective target density (3\,\%).\par
For the lowest energy points of refs. \cite{Formicola04-PLB,Imbriani05-EPJA}, not all transitions given in table \ref{Motivation-table} were measured, and R-matrix extra\-polations \cite{Imbriani05-EPJA} had to be added, giving a 7\,\% contribution. An analogous procedure had to be applied to the lowest data points of ref. \cite{Runkle05-PRL},
with R-matrix fit results \cite{Runkle05-PRL,Imbriani05-EPJA} contributing 26\,\% to the lowest data point. In the resulting total S-factor picture (fig. \ref{Sfactor.eps}), the present data are systematically lower than ref. \cite{Lamb57-PR}, whereas good agreement is obtained with ref. \cite{Schroeder87-NPA} in the overlapping energy region.
Excellent agreement is obtained between the present data and refs. \cite{Formicola04-PLB,Imbriani05-EPJA}. Good agreement is observed with ref. \cite{Runkle05-PRL}, except for $E$\,=\,185\,--\,215\,keV, where ref. \cite{Runkle05-PRL} is systematically lower than both the present data and refs. \cite{Formicola04-PLB,Imbriani05-EPJA}.\par

In addition to the cross section measurement, the strength $\omega\gamma$ of the 259\,keV resonance has been determined: $\omega\gamma$ = 12.8$\pm$0.3$_{\rm stat}$$\pm$0.4$_{\rm syst}$ meV, in excellent agreement with previous works: 12.9$\pm$0.4$_{\rm stat}$$\pm$0.8$_{\rm syst}$\,meV \cite{Imbriani05-EPJA},
13.5$\pm$1.2\,meV \cite{Runkle05-PRL} and 14$\pm$1\,meV \cite{Becker82-ZPA}. The value by ref. \cite{Becker82-ZPA} has 
% also 
been adopted by the Nuclear Astrophysics Compilation of Reaction Rates (NACRE) \cite{NACRE99-NPA}.\par

\begin{figure*}[p]
 \centering 
 \includegraphics[width=\figurewidth]{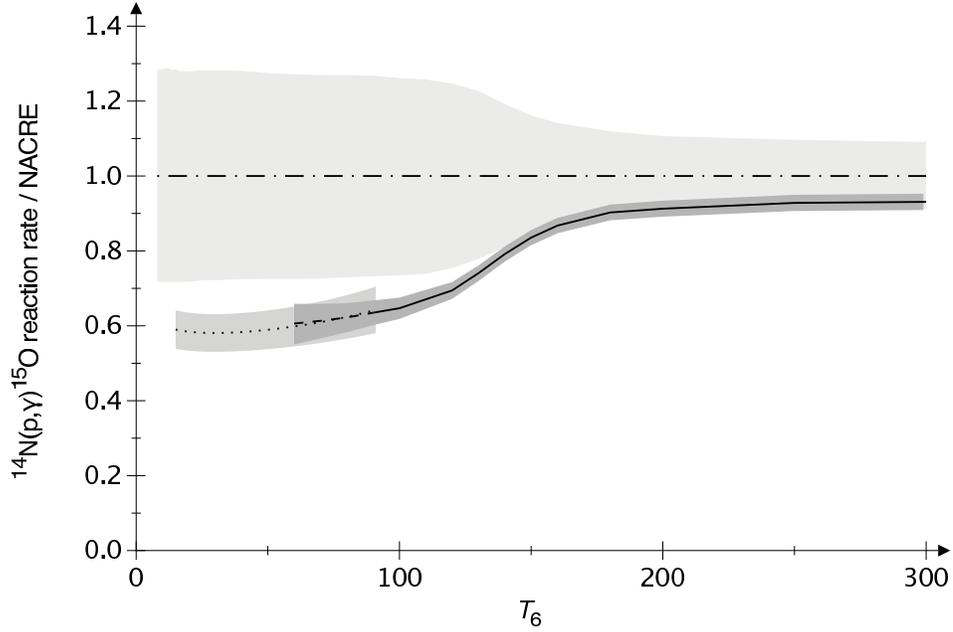}
 \caption{Thermonuclear reaction rate relative to the NACRE \cite{NACRE99-NPA} rate. Dot-dashed line: NACRE rate. Solid (dashed) line: present work and more than 90\,\% (more than 50\,\%) of the Gamow peak covered by experimental data. Dotted line: Extrapolation-based rate from ref. \cite{Imbriani05-EPJA}. The shaded areas indicate quoted upper and lower limit for the NACRE rate \cite{NACRE99-NPA} and $\pm$1$\sigma$ statistical uncertainty for the rate from ref. \cite{Imbriani05-EPJA} and the present work. }
 \label{Rate.eps}
\end{figure*}

The present astrophysical S-factors, corrected for the electron screening effect \cite{Assenbaum87-ZPA} in the adiabatic limit (10\,\% correction for $E$\,=\,70\,keV, 3\,\% for $E$\,=\,150\,keV), and the present $\omega\gamma$ value have been used to compute the thermonuclear reaction rate (fig. \ref{Rate.eps}). For energies $E$\,$<$\,70\,keV, the S-factor has been assumed to be constant and equal to the 70\,keV value. 
For temperatures $T_6$\,$\geq$\,90, the experimental data from the present work account for more than 90\,\% of the area under the Gamow peak, with the remaining 10\,\% depending on the assumption made for $E$\,$<$\,70\,keV. 
For 90\,$>$\,$T_6$\,$\geq$\,60, data account for 50 -- 90\,\% of the area under the Gamow peak. \par

For $T_6$\,$>$\,180, the rate is dominated by the 259\,keV resonance, and the present, lower resonance strength leads to a rate that is systematically 10\,\% lower than NACRE. 
For $T_6$\,$<$\,180, nonresonant capture becomes more and more important, and the present, experiment-based rate is up to 40\,\% lower than NACRE. \par

% =========================================================
\section{Astrophysical consequences}
% =========================================================

In several astrophysical scenarios the present, revised reaction rate for $T_6$\,$\geq$\,60 has direct consequences. After the end of their helium burning phase, low-mass stars burn hydrogen and helium, respectively, in two shells surrounding a degenerate carbon--oxygen core and give rise to the so-called asymptotic giant branch (AGB) \cite{Iben83-ARAA} in the Hertzsprung-Russell diagram. Flashes of helium burning spawn convective mixing in a process called dredge-up, transporting the products of nuclear burning from inner regions of the star to its surface, where they are in principle accessible to astronomical observations.
The temperature in the hydrogen burning shell is of the order of $T_6$\,=\,50\,--\,80 for a 2\,$M_\odot$ AGB star ($M_\odot$: mass of the Sun) with metallicity $Z$\,=\,0.01 \cite{Herwig04-ApJ}. 
It has been shown that an arbitrary 25\,\% reduction of the \nuc{14}{N}(p,$\gamma$)\nuc{15}{O} rate with respect to NACRE leads to twice as efficient dredge-up of carbon to the surface of the star \cite{Herwig04-ApJ}. At these temperatures, the present rate is 40\,\% below NACRE (fig. \ref{Rate.eps}). 
Recently, a simulation for a 5\,$M_\odot$, $Z$\,=\,0.02 AGB star \cite{Weiss05-AA} found stronger thermal flashes for a reduced CNO rate, consistent with the result by ref. \cite{Herwig04-ApJ}. Explosive burning in novae \cite{Jose98-ApJ} takes place at even higher temperatures, typically $T_6$\,$\approx$\,200, through the hot CNO cycle. The abundance of \nuc{15}{N} (daughter of \nuc{15}{O}) in nova ashes depends sensitively on the \nuc{14}{N}(p,$\gamma$)\nuc{15}{O} rate \cite{Iliadis02-ApJSS}; the present, significantly more precise rate reduces the nuclear uncertainty of the predicted abundance. \par

% =========================================================
\section{Summary}
% =========================================================

The total cross section of the \nuc{14}{N}(p,$\gamma$)\nuc{15}{O} reaction, the bottleneck of the CNO cycle, has been measured for $E$\,=\,70\,--\,228\,keV, with typically 3\,\% (at most 10\,\%) statistical and 5\,\% (at most 7\,\%) systematic uncertainty. For the first time, precision cross section data has been obtained directly at energies of hydrogen burning in AGB stars. The strength of the $E_{\rm R}$\,=\,259\,keV resonance has been determined with improved precision. The thermonuclear reaction rate for several scenarios of stable and explosive hydrogen burning has been calculated directly from the present experimental data, with negligible impact from assumptions made for lower energies. Several significant consequences of the present, experiment-based rate for the evolution of AGB stars and for nucleosynthesis in novae have been discussed. \par

% =========================================================
% Acknowledgements
% =========================================================

\begin{ack}
The authors are indebted to the INFN technical staff at Gran Sasso and in Genova, in particular to P.\,Cocconi and F.\,Parodi, for their support during the experiment. 
---
This work was supported by INFN and in part by: TARI RII-CT-2004-506222, OTKA T 42733 and T 49245, and BMBF (05CL1PC1-1).
\end{ack}

% =========================================================

\clearpage

% ================== 
\begin{table*}[h]
\centering
\renewcommand{\arraystretch}{0.9}
\caption{Cross section measurements of \nuc{14}{N}(p,$\gamma$)\nuc{15}{O} at low energy. Capture to the states at 6.859 and 7.276\,MeV in \nuc{15}{O} is negligible \cite{Schroeder87-NPA,Mukhamedzhanov03-PRC}. Uncertainties given refer to the lowest energy point. $T_6^{\,\rm min}$ is the lowest temperature where more than 90\,\% of the integral over the Gamow peak is covered by experimental data.}
\label{Motivation-table}
\begin{tabular}{|l|c|lllll|c|rr|c|}
\hline
Year & Ref. & \multicolumn{5}{c|}{Data on capture to states:} & $E$ [keV] & \multicolumn{2}{c|}{Uncertainty} &  $T_6^{\,\rm min}$ \\
 & & & & & & & & stat. & syst. & \\ \hline
1957 & \cite{Lamb57-PR} & \multicolumn{5}{c|}{All (activity measurement)} & 93 -- 126 & 46\,\% & 15\,\% & none\footnotemark \\
1987 & \cite{Schroeder87-NPA} & GS & 5.181 & 5.241 & 6.172 & 6.792  & 181 -- 3600 & 32\,\% & 13\,\% & 180\\
2004/5 & \cite{Formicola04-PLB,Imbriani05-EPJA} & GS & 5.181 & 5.241 & 6.172 & 6.792 & 119 -- 367 & 7\,\% & 10\,\% & 150 \\
2005 & \cite{Runkle05-PRL} & GS & & & 6.172 & 6.792 & 134 -- 482 & 38\,\% & 9.4\,\% & 170 \\ 
\hline \hline
\multicolumn{2}{|l|}{Present work} & \multicolumn{5}{c|}{All (4$\pi$ summing crystal)} & 70 -- 228, 259 & 10\,\% & 7\,\% & 90 \\
\hline
\end{tabular}
\end{table*}
% ================== 
%
\footnotetext{The energy range of ref. \cite{Lamb57-PR} is so narrow that it does not cover more than 50\,\% of the Gamow peak for any given temperature.}

\clearpage

% ============ Spectrum.eps
% ============

% ============ Sfactor.eps

% ============ Rate.eps
% ============ 

\end{document}